\documentclass[doublespacing]{elsart}

\usepackage{amssymb}
\usepackage{graphicx}
\usepackage{amsmath}

\journal{}

\begin{document}

\begin{frontmatter}

\title{On the equilibrium of a charged massive particle in the field of a Reissner-Nordstr\"om  black hole}

\author[iac,icra]{D. Bini},
\ead{binid@icra.it}
\author[icra]{A. Geralico},
\ead{geralico@icra.it}
\author[icra,fisica]{R. Ruffini}
\ead{ruffini@icra.it}

\address[iac]
{Istituto per le Applicazioni del Calcolo ``M. Picone'', CNR I-00161 Rome, Italy}

\address[icra]
{International Center for Relativistic Astrophysics - I.C.R.A.\\
University of Rome ``La Sapienza'', I-00185 Rome, Italy}

\address[fisica]
{Physics Department, University of Rome ``La Sapienza'', I-00185 Rome, Italy}

\begin{abstract}
The multiyear problem of a two-body system consisting of a Reissner-Nordstr\"om black hole and a charged massive particle at rest is here solved by an exact perturbative solution of the full Einstein-Maxwell system of equations. 
The expressions of the metric and of the electromagnetic field, including the effects of the electromagnetically induced gravitational perturbation and of the gravitationally induced electromagnetic perturbation, are presented in closed analytic formulas. 
\end{abstract}


\end{frontmatter}

The study of a massive charged particle in equilibrium in a Reissner-Nordstr\"om black hole background presents a variety of conceptual issues still widely open after more than twenty years of research, ranging from the classical aspects of general relativity to the quantum aspects of black hole tunneling processes (see e.g. Parikh and Wilczek \cite{PW}).

The problem of the interaction of a charged particle, neglecting its mass contribution, with a Reissner-Nordstr\"om black hole was addressed by Leaute and Linet \cite{leaute}.
They extended previous results obtained in the case of a Schwarzschild spacetime by Hanni \cite{hannijth}, Cohen and Wald \cite{CoW}, Hanni and Ruffini \cite{HR} and Linet \cite{linet} himself.
Their study was done in the test field approximation neglecting the backreaction both of the mass and of the charge of the particle on the background electromagnetic and gravitational fields.

We here approach the complete problem of a massive charged particle of mass $m$ and charge $q$ at rest in the field of a Reissner-Nordstr\"om black hole with mass $\mathcal{M}$ and charge $Q$. The full Einstein-Maxwell system of equations are solved taking into account the backreaction on the background fields due to the presence of the charged massive particle. 
The source terms of the Einstein equations contain the energy-momentum tensor associated with the particle's mass, the electromagnetic energy-momentum tensor associated with the background field as well as additional interaction terms, first order in $m$ and $q$.
Such terms are proportional to the product of the square of the charge $Q$ of the background geometry and the mass $m$ of the particle ($\sim Q^2m$) and to the product of the charges of both the particle and the black hole ($\sim qQ$). These terms originate from the \lq\lq electromagnetically induced gravitational perturbation'' \cite{jrz2}. 
On the other hand, the source terms of the Maxwell equations contain the electromagnetic current associated with the particle's charge as well as interaction terms proportional to the product of the black hole's charge $Q$ and the mass $m$ of the particle ($\sim Qm$), originating the \lq\lq gravitationally induced electromagnetic perturbation'' \cite{jrz1}.  

We summarize here the main results based on the first order perturbation approach formulated by Zerilli \cite{Zerilli} using the tensor harmonic expansion of both the gravitational and electromagnetic fields.
Details will be found in \cite{bgrPRD}.

The Reissner-Nordstr\"om black hole metric is given by
\begin{eqnarray}
\label{RNmetric}
ds^2&=&- f(r)dt^2 + f(r)^{-1}dr^2+r^2(d\theta^2 +\sin ^2\theta d\phi^2)\ ,
\nonumber\\
f(r)&=&1 - \frac {2\mathcal{M}}{r}+\frac{Q^2}{r^2}\ ,
\end{eqnarray}
with associated  electromagnetic field
\begin{equation}
\label{RNemfield}
F=-\frac{Q}{r^2}dt\wedge dr\ ,
\end{equation}
and the horizon radii are $r_\pm={\mathcal M}\pm\sqrt{{\mathcal M}^2-Q^2}={\mathcal M}\pm\Gamma$. 
We consider the case $ |Q|\leq {\mathcal M} $ and the region $r>r_+$ outside the outer horizon.
The \lq\lq extreme'' charged hole corresponds to $|Q|={\mathcal M}$.

The particle is assumed to be at rest at the point $r=b $ on the polar axis $\theta=0$. 
The only nonvanishing components of the stress-energy tensor and of the current density are given by
\begin{eqnarray}
\label{sorgenti}
T_{{00}}^{\rm{part}}&=&{\frac {m}{2\pi {b}^{2}}}f(b)^{3/2}\delta  \left( r-b \right) \delta  \left( \cos \theta -1 \right)\ ,\nonumber\\
J^{{0}}_{\rm{part}}&=&{\frac {q}{2\pi {b}^{2}}}\delta  \left( r-b \right) \delta  \left( \cos  \theta -1 \right)\ ,
\end{eqnarray}
and the combined Einstein-Maxwell equations are thus
\begin{eqnarray}
\label{EinMaxeqs}
\tilde G_{\mu \nu }&=&8\pi \left(T_{\mu \nu }^{\rm{part}} + \tilde T_{\mu \nu }^{\rm em}\right)\ ,\nonumber\\
\tilde F^{\mu \nu }{}_{;\,\nu }&=& 4\pi  J^{\mu }_{\rm{part}}\ , \quad {}^* \tilde F^{\alpha\beta}{}_{;\beta}=0\ .
\end{eqnarray}
The quantities denoted by a tilde refer to the total electromagnetic and gravitational fields, to first order of the perturbation
\begin{eqnarray}
\label{pertrelations}
\tilde g_{\mu \nu }&=&g_{\mu \nu } + h_{\mu \nu }\ ,\nonumber\\
\tilde F_{\mu \nu }&=&F_{\mu \nu }+  f_{\mu \nu }\ ,\nonumber\\
\tilde T_{\mu \nu }^{\rm em}&=&\frac1{4\pi}\left[\tilde g^{\rho \sigma }\tilde F_{\rho \mu }\tilde F_{\sigma \nu } - \frac14\tilde g_{\mu \nu }\tilde F_{\rho \sigma }\tilde F^{\rho \sigma }\right]\ ,\nonumber\\
\tilde G_{\mu \nu }&=&\tilde R_{\mu \nu }-\frac12\tilde g_{\mu \nu }\tilde R\ .
\end{eqnarray}
Note that the covariant derivative operation makes use of the perturbed metric $\tilde g_{\mu \nu }$.
The corresponding quantities without a tilde refer to the background Reissner-Nordstr\"om metric (\ref{RNmetric}) and electromagnetic field (\ref{RNemfield}).
Following Zerilli's \cite{Zerilli} procedure we expand the fields $h_{\mu \nu }$ and $f_{\mu \nu }$ as well as the source terms (\ref{sorgenti}) in tensor and scalar harmonics respectively (see Tables I, II, III and V in Ref. \cite{Zerilli}). The perturbation equations are then obtained from the system (\ref{EinMaxeqs}), keeping terms to first order in the mass $m$ of the particle and its charge $q$.
The axial symmetry of the problem about the $z$ axis ($\theta=0$) allows to put the azimuthal parameter equal to zero in the expansion, leading to a great simplification.
Furthermore, it is sufficient to consider only electric-parity perturbations, since there are no magnetic sources \cite{jrz2,jrz1,Zerilli}.  

The Regge-Wheeler gauge \cite{ReggeW} leads to the following set of gravitational and electromagnetic perturbation functions: $K$, $H_0$, $H_2$, $\tilde f_{{01}} $ and $\tilde f_{{02}}$.
The Einstein-Maxwell field equations (\ref{EinMaxeqs}) give rise to the following system of radial equations for values $l\geq2$ of the multipoles 
\begin{eqnarray}
  \label{eq1RN}  
0&=&{e^{2\nu}} \left[ 2K{}''-\frac2rW{}'+\left(\nu{}'+\frac6r\right) { K{}'}-    4\left(\frac1{r^2}+\frac{\nu{}'}r\right)W \right]-\frac{2\lambda e^{\nu}}{r^2}(W+K)\nonumber\\
&& -2{\frac {{Q}^{2}{e^{\nu}}W }{{r}^{4}}}-4{\frac{Q{e^{\nu}}\tilde     f_{{01}}}{{r}^{2}}}+A_{{00}}\ , \nonumber\\
  \label{eq2RN} 
0&=&\frac 2rW{}'-\left(\nu{}'+\frac2r\right)K{}' -\frac{2\lambda e^{-\nu}}{r^2}(W-K)-        2{\frac{{Q}^{2}{e^{-\nu}}W }{{r}^{4}}}+4{\frac {Q{e^{-\nu}}\tilde f_{{01}} }{{r}^{2}}}\ ,\nonumber\\
  \label{eq3RN} 
0&=&K{}''+\left(\nu{}'+\frac2r\right)K{}'- W{}'' -2\left(\nu{}'+\frac1r\right)W{}'\nonumber\\
&&+\left(\nu{}''+{\nu{}'}^2+\frac{2\nu{}'}r\right)(K-W) -2{\frac{{Q}^{2}{e^{-\nu}}K}{{r}^{4}}}
    +\frac{4Q{e^{-\nu}}}{{r}^{2}}  \tilde f_{{01}}\ ,\nonumber\\
  \label{eq4RN} 
0&=&-W{}' + K{}' -\nu{}' W +4{\frac {Q{e^{-\nu}}\tilde f_{{02}} }{{r}^{2}}}\ ,\nonumber\\
  \label{eq5RN} 
0&=&\tilde f_{{01}}{}'+\frac2{r}\tilde f_{{01}} -{\frac {l\left( l+1 \right) {e^{-\nu}}\tilde     f_{{02}} }{{r}^{2}}}-{\frac {Q}{{r}^{2}}}K{}' +4\pi v\ ,\nonumber\\
  \label{eq6RN} 
0&=&\tilde f_{{01}} - \tilde f_{{02}}{}'\ ,
\end{eqnarray}
where $\lambda=\frac12 \left( l-1 \right) \left( l+2 \right)$, $H_0= H_2\equiv W$ and $e^{\nu}=f(r)$ is Zerilli's notation; a prime denotes differentiation with respect to $r$.
The quantities 
\begin{equation}
\label{sorgexp}
A_{{00}}=8\sqrt{\pi} \frac{m\sqrt {2l+1}}{b^2}f(b)^{3/2}\delta \left( r-b \right)\ , \qquad
v=\frac1{2\sqrt{\pi}} \frac{q\sqrt {2l+1}}{b^2}\delta \left( r-b \right)\ 
\end{equation}
come from the expansion of the source terms (\ref{sorgenti}).

We have a system of $6$ coupled ordinary differential equations for $4$ unknown functions: $K$, $W$, $\tilde f_{{01}} $ and $\tilde f_{{02}}$.
The compatibility of the system requires that the following stability condition holds
\begin{equation}
\label{bonnoreqcond}
m=qQ\frac{b f(b)^{1/2}}{{\mathcal M}b-Q^2}\ ,
\end{equation}  
involving the black hole and particle parameters as well as their separation distance $b$.
This condition coincides with the equilibrium condition for a test particle of mass $m$ and charge $q$ in the field of a Reissner-Nordstr\"om black hole given by Bonnor \cite{bonnor}. 
There he simply considered the classical expression for the equation of motion of the particle 
\begin{equation}
m U^\alpha \nabla_\alpha U^\beta =q F^\beta{}_\mu U^\mu\ ,
\end{equation} 
with 4-velocity $U =f(r)^{-1/2}\partial_t$, neglecting all the feedback terms, and obtained Eq. (\ref{bonnoreqcond}) as the equilibrium condition. 
The coincidence of these results is quite surprising, since our gravitational and electromagnetic fields including all the feedback terms are quite different from those used by Bonnor.

If the black hole is \lq\lq extreme'', then from Eq. (\ref{bonnoreqcond}) follows that also the particle must have $q/m=1$, and equilibrium exists independent of the separation.
In the general non-extreme case $Q/\mathcal{M}<1$ there is instead only one position of the particle which corresponds to equilibrium, for any given value of the charge-to-mass ratios of the bodies. In this case the particle charge-to-mass ratio must satisfy the condition $q/m>1$. 

We now give the general expression for both the perturbed gravitational and electromagnetic fields in closed analytic form by summing over all multipoles of the Zerilli expansion \cite{bgrPRD}.
The perturbed metric is given by
\begin{equation}
\label{lineelemnonextr}
d{\tilde s}^2=-[1-{\bar {\mathcal H}}]f(r)dt^2 + [1+{\bar {\mathcal H}}][f(r)^{-1}dr^2+r^2(d\theta^2 +\sin ^2\theta d\phi^2)]\ ,
\end{equation}
where 
\begin{equation}
\label{barH}
{\bar {\mathcal H}}=2\frac{m}{br}f(b)^{-1/2}\frac{(r-{\mathcal M})(b-{\mathcal M}) -\Gamma^2\cos\theta}{{\bar {\mathcal D}}}\ ,
\end{equation} 
with
\begin{equation}
\label{barD}
{\bar {\mathcal D}} = [(r-{\mathcal M})^2+(b-{\mathcal M})^2 - 2(r -{\mathcal M})(b-{\mathcal M})\cos\theta- \Gamma^2\sin^2\theta]^{1/2}\ . 
\end{equation}
Note that in the extreme case $Q/\mathcal{M}=q/m=1$ this solution reduces to the linearized form of the well known exact solution by Majumdar and Papapetrou \cite{maj,pap} for two extreme Reissner-Nordstr\"om black holes.
The asymptotic mass measured at large distances by the Schwarzschild-like behaviour of the metric of the whole system consisting of black hole and particle is given by
\begin{equation}
M_{\rm eff}={\mathcal M}+m+E_{\rm int}\ ,
\end{equation} 
where the interaction energy turns out to be
\begin{equation}
E_{\rm int}=-m\left[1-\left(1-\frac{{\mathcal M}}{b}\right)f(b)^{-1/2}\right]\ .
\end{equation} 
It can be shown that this perturbed metric is spatially conformally flat; moreover, the solution remains valid as long as the condition $|{\bar {\mathcal H}}|\ll1$ is satisfied.

The nonvanishing components of the perturbed electric field are given by
\begin{eqnarray} 
\label{Zeremtensorpertnonextr} 
E_r&=&\frac{q}{r^3}\frac{{\mathcal M}r-Q^2}{{\mathcal M}b-Q^2}\frac1{{\bar {\mathcal D}}}\bigg\{-\bigg[{\mathcal M}(b-{\mathcal M})+\Gamma^2\cos\theta\nonumber\\
&&+[(r-{\mathcal M})(b-{\mathcal M})-\Gamma^2\cos\theta]\frac{Q^2}{{\mathcal M}r-Q^2}\bigg]\nonumber\\
&&+\frac{r[(r-{\mathcal M})(b-{\mathcal M})-\Gamma^2\cos\theta]}{{\bar {\mathcal D}}^2}[(r-{\mathcal M}) -(b-{\mathcal M})\cos\theta]\bigg\}\ ,
\nonumber\\ 
E_{\theta}&=&q\frac{{\mathcal M}r-Q^2}{{\mathcal M}b-Q^2}
\frac{b^2f(b)f(r)}{{\bar {\mathcal D}}^3}\sin\theta\ . 
\end{eqnarray} 
The total electromagnetic field to first order of the perturbation is then
\begin{equation}
\label{RNemfieldpertnonextr}
\tilde F=-\left[\frac{Q}{r^2}+E_r\right]dt\wedge dr - E_{\theta}dt\wedge d\theta\ .
\end{equation}
The total perturbed electrostatic potential is given by
\begin{equation}
\label{relperttest}
V_{\rm tot}=V_{\rm test}+\left[1+\frac12\left(1-\frac{r}b\right){\bar {\mathcal H}}+\frac{qQ}{{\mathcal M}b-Q^2}\left(1-\frac{{\mathcal M}}b\right)\right]V^{\rm{BH}}\ , 
\end{equation}
where $V^{\rm{BH}}=Q/r$ is the black hole electrostatic potential, while $V_{\rm test}$ denotes the electrostatic potential of the particle obtained within the test-field approximation by Leaute and Linet \cite{leaute}
\begin{equation}
\label{solRNpot}
V_{\rm test} = \frac q{b r} \frac{(r-{\mathcal M})(b-{\mathcal M})
 -\Gamma^2\cos\theta}{{\bar {\mathcal D}}} + \frac{q{\mathcal M}}{b r}\ .
\end{equation}
The second and third terms in the bracketed expression of (\ref{relperttest}) represent the \lq\lq gravitationally induced'' and \lq\lq electromagnetically induced'' electrostatic potential respectively and the equilibrium condition (\ref{bonnoreqcond}) has been conveniently used.

The Zerilli's procedure of expansion of both the gravitational and electromagnetic fields in tensor harmonics 
is largely used in the literature to study linear perturbations of spherically symmetric spacetimes due to 
some external source. 
We have given here the analytic solution for a problem which has raised much interest and discussions for many years.
We have obtained closed form expressions for both perturbed metric and electromagnetic field 
due to a charged massive particle at rest in the field of a Reissner-Nordstr\"om  black hole, taking
advantage of the static character of the perturbation as well as of the axial symmetry of the configuration.
The infinite sum of multipoles converges to an analytic form. 

In addition to its theoretical significance, this result can become an important tool in testing the validity of numerical investigations addressing the dynamics of many body solutions in Einstein-Maxwell systems.

\end{document}